# Techno-economic analyses for vertical use cases in the 5G domain

Sandrine Destouet Roblot[1], Mythri Hunukumbure[2], Nadege Varsier[1], Elena Serna Santiago[3], Yu Bao[1], Serge Langouet[1], Marie-Hélène Hamon[1], Sebastien Jeux[1]

[1] Orange Labs, 4 rue du clos courtel 35512 Cesson-Sévigné, France
[2] Samsung R&D Institute UK, Communications House, Staines-upon-Thames, UK. TW18 4QE
[3] Telefonica I+D, Ronda de la Comunicación, s/n, 28050 Madrid, Spain

*Abstract*— **This paper provides techno-economic analyses on the network deployments to cover 4 key verticals, under 5G-NR. These verticals, namely Automotive, Smart city, Long range connectivity and Disaster and emergency support, were chosen to reflect the ONE5G project objective of investigating environments from densely populated cities ("Megacity") to large underserved areas. The work presented covers the network deployment framework including common centralization strategies and the main cost factors. Initial results presented for long range connectivity and emergency support networks provide the cost trade-offs in different deployment options and cost sensitivity to some of the parameters.**

*Keywords—5G, cost models, V2X, long range, smart cities, underserved areas, TCO Introduction*

## I. INTRODUCTION

The foreseen explosion in new digital services, new vertical markets and diverse applications will make 5G networks unique in many aspects. The techno-economic analysis of such complex networks is of greatest importance, as it assesses the economic viability of new services, especially in vertical sectors not addressed by cellular networks before. In 5G-PPP project ONE5G [1] in particular, the aim of developing a flexible air-interface able to be efficient as well as in dense urban environments (labeled "Megacities" in ONE5G) as in underserved areas scenarios [2], comes with the objective of identifying the cost driving elements for the roll-out and operation of systems in such scenarios.

Four use cases (uc) were selected from a total of 9, to be studied in ONE5G techno-economic analysis: uc1 on assisted, cooperative and tele-operated driving, uc3 on smart cities, uc4 on long range connectivity in remote areas and uc9 on Ad-hoc airborne platforms for disasters and emergencies. This choice was realized with the aim of fairly reflecting the Megacities and Underserved areas scenarios as well as the 3 service categories targeted by 5G (e.g. eMBB, URLLC and mMTC).

Multiple deployment options were also considered, either based on 3GPP 5G New Radio (NR) Rel.15 network and accounting for the technical needs of each use case. Technology selection (such as Multi-access Edge Computing for uc1 or cellular IoT for uc3) was governed by the primary requirements of each uc and the cost impacts of different methodologies to meet the same requirements were studied.

## II. METHODOLOGY

The starting point of the techno-economic study is to consider an already existing NR 5G Rel.15 network as defined in 3GPP Rel.15 [3]. It follows a hybrid model where the gNB can be either aggregated in a single node or disaggregated into three logical nodes comprising the Remote Radio Head (RRH), the distributed unit (DU) and the centralized unit (CU). These centralization themes are a key common factor in uc1, uc4 and uc9 analysis, where the 3GPP centralization options 2 (PDCP layer) and 7 (upper PHY layer) are considered.

The related backhaul and fronthaul (last part of the transport network that reaches the base station or Radio Unit - plus the metro segment between the Baseband Unit (BBU) and the RRH) cost models are developed in line with the work performed in mmMAGIC project [4]. Different possibilities can be envisioned for the backhaul and fronthaul deployments, based on owned microwave, leased or owned fiber. In all cases the cost is dependent of the fronthaul or backhaul capacity. For the last drop, either leased lines or owned lines are envisioned: in leased lines, there is no need to consider a particular cost model as such cost would be subsumed within the overall connection cost charged by the third party that provides the connectivity; in owned lines case this cost is calculated separately including civil work costs, average last drop length, and related equipment.

## III. TECHNO-ECONOMIC ANALYSIS

### A. Use case 1: assisted, cooperative and tele-operated driving

The assisted, cooperative and tele-operated driving *uc* is one of the multiple services that are comprised within V2X category that can pose stringent requirements over networks due to the sensitivity of its actions.

These requirements (low latency, high availability and very high reliability) are quite difficult to fulfill by traditional networks, requiring in most cases the need to deploy additional network nodes to satisfy the ultra-low communications that are needed. These nodes are known as Multi-access Edge Computing (MEC) nodes.

The techno-economic analysis carried out in *uc1* considers precisely this approach where one or several MEC nodes are included under different RAN topologies (Centralized (C-RAN) and Distributed (D-RAN)) [5] with the aim to reduce the E2E service latency and improve reliability. Megacities and underserved areas were considered in the study as both scenarios can seamlessly provide V2X services regardless their characteristics and main differences. The study was conducted in a two-step analysis.

Firstly, we analyzed the impact of a MEC node over the existing infrastructure in a D-RAN/C-RAN topology such as the power and fronthaul/backhaul requirements and how they should scale up/down with the MEC introduction. Lastly, the cost impact of deploying V2X with the aid of MEC nodes for such different scenarios configurations was quantified.

The initial outcomes of the quantitative analysis show in terms of Total Cost of Ownership (TCO) that C-RAN split 7 (Phy layer) is the most cost-effective solution to provide V2X services regardless the characteristics of the scenario, i.e. number of sectors, MIMO order, modulation scheme and bandwidth, in both megacities (Figure 1) and rural areas (Figure 2). D-RAN remains as the most expensive option due to the high dependency on dedicated hardware, whereas C-RAN split 2 (PDCP layer) architecture slightly reduces the total cost as it starts avoiding this dependency by introducing Commercial Of The Shelf (COTS) equipment.

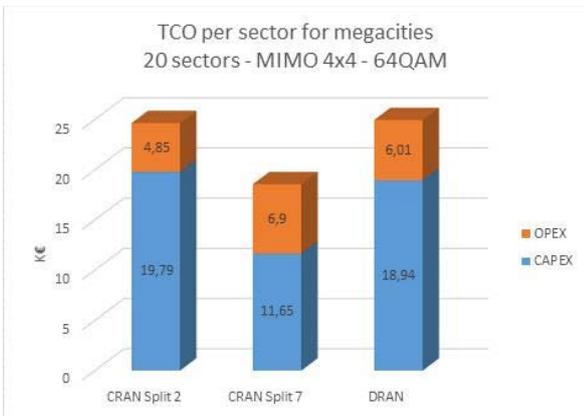

**Figure 1: TCO per sector for Megacity scenario**

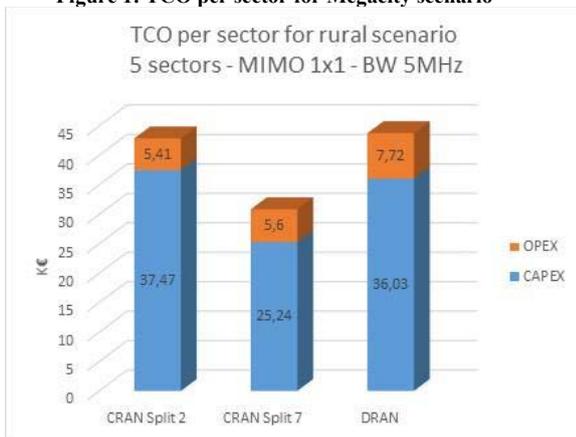

**Figure 2: TCO per sector for Underserved area scenario**

### B. Use case 3: smartcities

The smart cities *uc* deals with non-time-critical processes and logistics for dense urban and suburban area management. The objective is to provide mMTC services for applications such as traffic management, waste collection and management, parking detection and information, air monitoring, etc. where data has small payloads and no high constraints on latency.

3GPP announced that such Low-Power Wide-Area (LPWA) deployments requirements would finally be supported by Narrowband-IoT (NB-IoT) and LTE-M whose evolution would be part of 5G NR, meaning that NB-IoT and LTE-M are on the path to 5G [6]. That's why the techno-economic analysis for use case 3 was considered with, on one side, the in-band deployment of LTE-M and of NB-IoT on the other side. The study was carried out by firstly evaluating the number of additional Physical Resource Blocks (PRBs) that would be necessary in Rel. 16 to satisfy the number of devices envisioned for the Smart cities applications; and secondly by assessing the impact on cost deployment. For that purpose, performance studies carried out for 3GPP by Ericsson were considered [7]. In order to see if LTE-M and NB-IoT would meet the 5G requirements in term of density of connections (1 million devices/km$^2$) Ericsson carried out non-full buffer system simulations considering two different inter-site distances (ISDs) and two different channel models. For the traffic model, devices were considered to emit 32 bytes messages every 2 hours. For NB-IoT (LTE-M respectively) the results have shown that 1 to 15 PRBs (1 to 3 narrowbands respectively) would be necessary to fulfil the 5G mMTC requirements, depending on the considered ISD and channel model. Then to evaluate the number of devices envisioned for the smart cities applications at the time of Rel.15 and Rel.16 deployments, a study was carried out extrapolating these numbers for a typical dense urban city (Paris) in 2020 (corresponding to Rel. 15) and 2030 (corresponding to Rel. 16):250 000 devices were extrapolated for Rel. 15 and 650 000 for Rel. 16.

The analysis has shown that, depending on the deployment settings and the traffic model, up to 6 additional PRBs (one additional narrowband respectively) would be necessary for NB-IoT (LTE-M respectively) between Rel. 15 and Rel. 16 to satisfy the number of devices envisioned for Smart cities applications as illustrated on Figure 3.

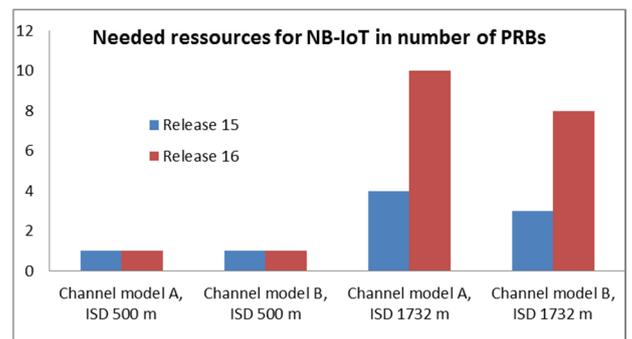

(a)

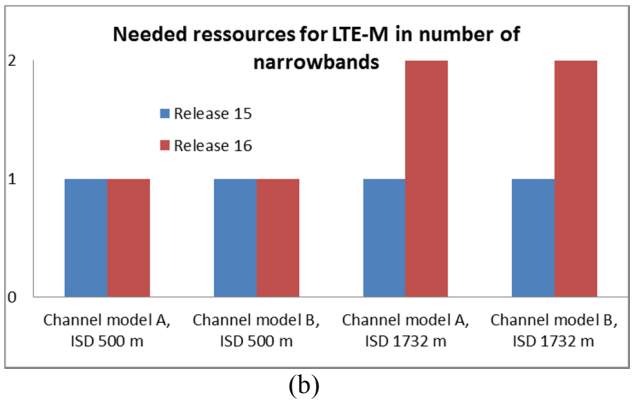

Figure 3: Needed resources for NB-IoT (respectively LTE-M) to satisfy to the number of devices envisioned for Smart cities applications in Rel. 15 and 16

*C. Use case 4: long range connectivity in remote areas*

The long range connectivity in remote areas *uc* is dedicated to underserved areas, for the provision of minimal voice and data services over long distances in low user density areas. The applications targeted are minimal services including voice over long distances plus best effort data services for smartphones, tablets, etc. The priority of this service is to provide a maximum coverage (up to 50 km in rural and 100 km or more for ultra-rural) without strict requirements on throughput.

Such wide radio coverage has some implications on the backhaul technology (fiber, microwave, satellite) and its configuration (architecture, topology). In both rural and extreme rural environments, different solutions are being analyzed to check which configuration guarantees the targeted coverage. This is performed by using specific link budget tools, long distance propagation model and 3GPP radio interface physical channels link budget data.

Multiple configurations were studied in order to determine the best alternative to meet the requirements of extreme rural and rural deployments: three antenna heights (60m, 75m, 108m), three options of vertical diversity (2, 3 or 4 floors), two MIMO orders (2x2 or 4x2), two possibilities of sectorization (3 or 6 sectors).

For extreme rural, the objective was to reach a 100km coverage with throughputs of 2 Mbps downlink and 0.256 Kbps uplink. The best option found was to consider the 700 MHz frequency band, a mast elevation of 108m, four floors of antennas, six sectors and MIMO4x2. With such configuration, the maximum coverage we obtained is 80km.

The extreme rural TCO per km² of such configuration is pictured in Figure 4. For this extreme rural case, the backhaul is supported by microwaves only. The satellite option was envisioned as well but is much more expensive and thus not considered as a possible solution, especially since we have wonders on its ability to maintain the latency requirements.

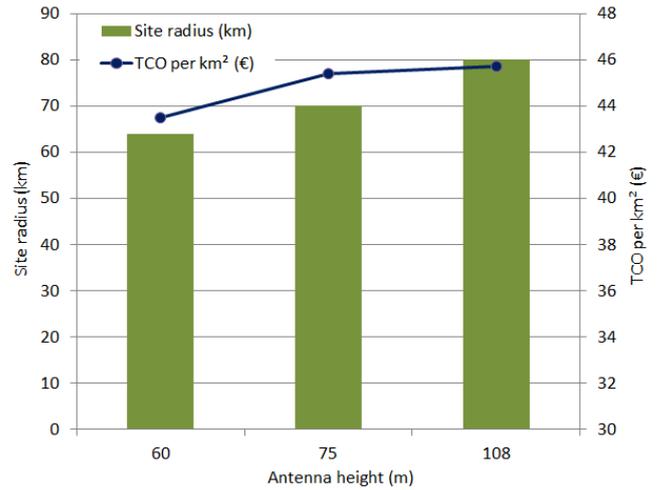

Figure 4: TCO per km² evolution with three different antenna heights, 6 sectors, MIMO4x2, 4 floors of antennas in extreme rural deployment

A similar analysis was performed for the rural areas case where the targeted objective was to reach 50km while maintaining throughputs of 50 Mbps downlink and 25 Mbps uplink. Massive MIMO at 3.5GHz was considered to fulfill the need for capacity: MIMO64x2 for downlink and SIMO1x64 for uplink. The best coverage we obtained was 15km.

Different options of backhaul were envisioned for the rural case, comparing the costs induced by C-RAN splits 2 and 7, mixing both microwave and dark fibers.

*D. Use case 9: Ad-hoc airborne platforms for disasters and emergencies*

This *uc* is customized to provide on demand, 5G eMBB services to emergency crews, wherever the need arises, within their service area. The 5G provision is through drones, which use multiple drone links to relay the signal (Fronthaul) back to a ground relay anchor station (existing 4G or 5G small cell upgraded to support this), then to the BBU and the core network as per the centralized (C-RAN) network architecture. The analysis considers 3 main deployment cost factors: the costs of acquiring and running a fleet of high precision drones and their radio kit, the costs of upgrading ground small cells and the incrementing costs of providing the additional fronthaul and backhaul capacity. For the fronthaul and backhaul, the leased line cost model is assumed, where the incremental costs are directly linked to the additional capacities. For the C-RAN split 2 (PDCP layer), the fronthaul capacity needed is lower than for the C-RAN split 7 (Phy layer). The capacity increment factors are taken from [4].

The overall TCO for this solution was estimated with the above 3 cost factors, for the two split points. The TCO – 1 year results are shown in Figure 5 below, where the split 2 costs are lower than for the split 7. The trade-off between higher drone costs and the lower fronthaul costs for the PDCP split outperforms the opposite trade-off for these factors in the upper PHY layer split. Similar results are also seen for the 5 year TCO.

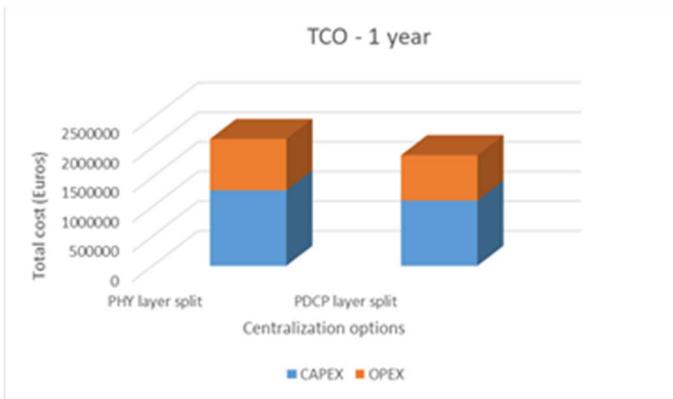

**Figure 5: TCO comparison for PHY and PDCP layer splits**

The cost sensitivity of the proposed solution to increments of the number of maximum drones per link is also studied. Incrementing the drone numbers increases the drone related costs, but reduces the number of ground small cells needing upgrades and the additional fronthaul capacities. Figure 6 shows that the TCO reduces exponentially with the drone numbers up to 6 and then increases again. The drone radio kit costs are taken here as to be comparable to ground small cells, but full results taking this also as a variable will be presented.

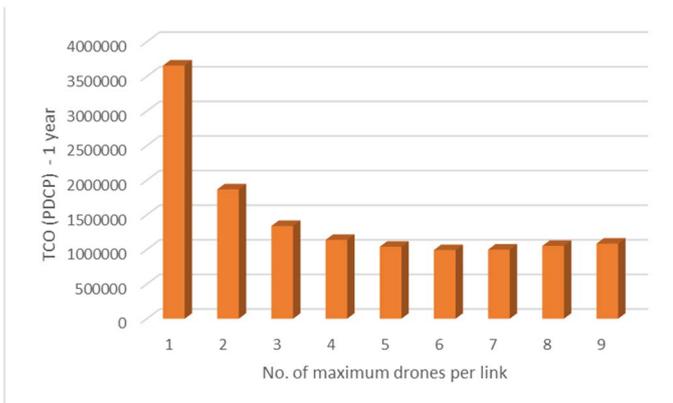

**Figure 6 TCO variation with the number of drones per link**

## IV. CONCLUSIONS

The presented work analyses the techno-economic aspects of 5G network deployment to support 4 key vertical areas. While the individual studies are quite distinct, common aspects such as 3GPP C-RAN options are utilized as a means of comparing the cost trends. Some of the cost models developed in the previous mmMAGIC project have been utilized to develop the studies. Some assumptions had to be made on the costs of certain network elements, as these envisaged 5G systems are still far from being deployed. So the trends we have highlighted are more accurate and the absolute cost values reported should be taken only as indicative

In the *uc1* (Automotive) analysis, the TCO for the C-RAN split 7 returns the lower costs than for the split 2 or the D-RAN options. This can be attributed to the hardware costs of the distributed units having more impact than the fronthaul capacity provision costs in *uc1*. In the *uc 9* (NTN emergency support) we see the reverse trend, where the distributed unit (drone) costs are lower than the Fronthaul and anchor BS upgrade costs.

For *uc3* we have shown that additional resources might be needed between Rel. 15 and Rel. 16 if we want to satisfy to the number of devices envisioned for Smart cities applications. This need will depend on the deployment settings and the considered traffic models.

In the *uc4* (long range), multiple deployment options were evaluated. Some parameters permit an improvement of the coverage while providing a more interesting TCO per km²; this is the case for the number of floors (vertical diversity) or the MIMO order. Sectorization increases the capacity while reducing the TCO per km² as well. On the other side, increasing the mast elevation of antennas enhances the coverage but does not reduce the TCO. The same way, increasing the power at the base station does not improve the coverage because the limiting factor is always on the uplink but has some negative impact on the TCO per km².

ACKNOWLEDGMENT

Part of this work has been performed in the framework of the Horizon 2020 project ONE5G (ICT-760809) receiving funds from the European Union. The authors would like to acknowledge the contributions of their colleagues in the project, although the views expressed in this contribution are those of the authors and do not necessarily represent the project.